\documentclass[12pt]{iopart}

\input{epsf}

\begin{document}

\title{Testing the Edwards hypothesis in spin systems under tapping
  dynamics}

\author{Johannes Berg$^{1}$, Silvio Franz$^{2,3}$ and Mauro
  Sellitto$^{2,4}$ }

\address{$^1$ Institut f\"{u}r Theoretische Physik, 
Universit\"{a}t zu K\"{o}ln, Z\"{u}lpicher Stra\ss e 77, 
D-50937 K\"{o}ln, Germany.}

\address{$^2$ The Abdus Salam International Centre for Theoretical
  Physics, Strada Costiera, 11 - 34100 Trieste, Italy.}
  
\address{$^3$ Service de Physique Th\'eorique, CEA Saclay, 91191 Gif
  sur Yvette cedex, France\footnote{Until December 31, 2001.}.}
  
\address{$^4$ Laboratoire de Physique, \'Ecole Normale Sup\'erieure de
  Lyon and CNRS, F-69007 Lyon, France\ddag.}

\begin{abstract}
  The Edwards hypothesis of ergodicity of blocked configurations for
  gently tapped granular materials is tested for abstract models of
  spin systems on random graphs and spin chains with kinetic
  constraints.  The tapping dynamics is modeled by considering two
  distinct mechanisms of energy injection: {\sl thermal} and {\sl
    random} tapping. We find that ergodicity depends upon the tapping
  procedure (i.e. the way the blocked configurations are dynamically
  accessed): for thermal tapping ergodicity is a good approximation,
  while it fails to describe the asymptotic stationary state reached
  by the random tapping dynamics.
\end{abstract}

\section{Introduction}

The probabilistic description of dissipative dynamical systems is a
central issue of modern statistical physics.  In general, the
non-equilibrium nature of the stationary state makes a general
principle analogous to the Boltzmann ergodic hypothesis for
Hamiltonian systems hard to find. The situation may be fortunate in
the case of gently tapped granular materials, where the dynamics
consists of cycles in which the system passes from a blocked
configuration to another through discrete injection of mechanical
energy (a tap) followed by a zero temperature relaxation under
gravity.  The observation of a reversible branch in the curve of the
asymptotic packing density versus the tapping amplitude, suggests the
existence of a stationary regime in which the packing density depends
monotonically on the vibration intensity~\cite{Chicago}.  On the other
hand, it is also known that macroscopic features of mechanically
stable packings (e.g. the packing density) depends on their collective
handling, i.e. the specific tapping procedure as borned out by
simulations~\cite{mehtabarker} and experiments~\cite{Pouliquen}. In
this situation one can ask for the invariant dynamical measure which
describes the sampling of the blocked configurations, and how it
depends on the energy injection mechanism.

The simplest hypothesis was made some times ago in a series of papers
by Edwards and co-workers where the uniform distribution over the
blocked states of given density was assumed independently of the
tapping procedure, provided it is extensive~\cite{Edo,cw}. This
proposal is particularly attractive as it leads by construction to a
thermodynamic framework analogous to that of ordinary thermal systems.
In particular, it leads to the concept of compactivity, which for
granular matter would play the same role of the temperature in
thermodynamic systems.

Effective temperatures also appear in the description of glassy
systems undergoing aging dynamics, which are by their nature far away
from their stationary state~\cite{CuKuPe}.  Their occurrence can be
justified by supposing that, although ergodicity does not hold at the
level of a single trajectory, trajectories corresponding to different
initial conditions and thermal histories sample finite life-time
states with asymptotically uniform measure~\cite{FrVi}.  The issue has
been investigated numerically with positive answer in 3D Lennard-Jones
glasses~\cite{SciTa}.  In the aging dynamics of a non-thermal kinetic
lattice-gas like the Kob-Andersen model~\cite{KoAn,KuPeSe}, the
generalised effective temperature~\cite{Se}, as well as more local
observables as the structure function, appears to be in agreement with
the corresponding ones computed from the Edwards measure on the
blocked states~\cite{BaKuLoSe}. Similar results have been obtained in
a realistic model of granular media under shear~\cite{Hernan}.
Further hints in favour of the Edwards hypothesis have been presented
in recent studies~\cite{Brey,bergmehta,Dean,LeDe,Annalisa,Vittoria} on
various kinds of spin models with tapping dynamics. These results
suggest a unified thermodynamic framework to describe aging glasses
and gently tapped granular systems~\cite{Jorge}.

Despite the fascination and the strong predictive power of a
statistical mechanical construction, it is at present not clear to
what extent and generality it actually applies to granular and glassy
systems, and what would they be the underlying reasons.  It is quite
natural in that context to look at the problem in abstract models,
which while mimicking the tapping energy injection mechanism and
subsequent dissipation, are easily amenable to numerical and
analytical investigations thus allowing to test the hypothesis in a
fine detail. Here we study two models of one dimensional kinetically
constrained systems (section 2), and some spin models on diluted
random graphs (section 3), by considering two distinct energy
injection mechanisms, which we call {\sl thermal} and {\sl random}
tapping.  We find the uniform measure to be a good approximation for
the thermal tapping, with improving accuracy at decreasing tapping
intensities.  While for the random tapping we observe systematic
deviations from the uniform measure for all finite tapping
intensities.  Moreover, in the case of kinetically constrained
systems, the validity of the approximation for thermal tapping
dynamics does not warrant its extension to the aging regime.  In the
case of random graph models we find that the validity of the uniform
measure also depends on whether or not there are neutral moves (single
spin flips which do not change the energy) and propose a modified
measure for the former case.

\section{Kinetically constrained spin chains}

The first two models we consider are a variant of the facilitated Ising
spin chains first introduced by Fredrickson and
Andersen~\cite{FrAn,Grant}, and its asymmetric version introduced by
J\"ackle and Eisinger~\cite{Jackle}.
These are abstract toy models of glassy behaviour whose  Hamiltonian 
is simply
\begin{equation}
  E = -\sum_{i=1}^N n_i \,,
  \label{H}
\end{equation}
where the $n_i=0,1$ are binary variables and the index $i$ runs over
the sites of a chain of length $N$ with periodic boundary condition.
Their dynamics is defined by the following kinetic constraints:
\begin{itemize}
  
\item {\bf Symmetric model (model S)} A variable can flip with a
  non-zero rate only if at least one of its neighbouring variable is
  equal to zero.  Specifically, the variables are randomly updated
  according to the transition matrix
  \begin{eqnarray}
    {\cal W}(n_i\to 1-n_i) & = &    
    \left( 1-n_{i+1} \, n_{i-1} \right)  
 \min \left[ 1, \,\exp(-\Delta E/T) \right] \,.
    \label{W_S}
  \end{eqnarray}
  
\item {\bf Asymmetric model (model A)} A variable can flip with a
  non-zero rate only if its {\sl left} neighbouring variable is equal
  to zero.  In this case the transition matrix is
  \begin{eqnarray}
    {\cal W}(n_i\to 1-n_i) & = &    
    \left( 1- n_{i-1} \right) 
	  \min \left[ 1, \,\exp(-\Delta E/T) \right] \,.
    \label{W_A}
  \end{eqnarray}

\end{itemize}
With these rules detailed balance is satisfied and the Markov chain
associated with the dynamic evolution at non-zero temperature is
irreducible on the full configuration space with the exception of the
configuration with the lowest energy (all the spins equal to one so 
the kinetic constraints prohibit a dynamical evolution). 
Therefore the approach to the canonic equilibrium distribution is
guaranteed.  However, after a quench at low temperature the density of
zeros become smaller and smaller and hence the relaxation become
sluggish as the kinetic constraints are hardly satisfied.  In spite of
their simple equilibrium measure, the finite temperature dynamics of
these models does not seem to be exactly solvable, but several
important results are 
known~\cite{Jackle,Reiter_sym,Schulz,Mauch_asym,SoEv,CrRiRoSe}. In
particular, the characteristic equilibration time at low-temperature
diverges as $\tau \sim {\rm e}^{b/T}$ for the model
S~\cite{Reiter_sym,Schulz}; and with a super-Arrhenius law $\tau
\sim{\rm e}^{a/T^2 }$, for the model A~\cite{Jackle,Mauch_asym,SoEv}.

One reason of special interest in these models is that they provide a
more severe test of the validity of the Edwards hypothesis since
they are characterised by the same entropy of blocked configurations
-- though their relaxational dynamics is qualitatively
different\footnote{There is actually a continuous class of dynamical
  models sharing the same entropy of blocked configurations,
  see~\cite{CrRiRoSe}.}.  Moreover, they offer the advantage that the
Edwards measure can be exactly computed and the analytic results
compared with the corresponding ones obtained from numerical
simulations of tapping.  In the following we will be mainly interested
in the stationary state reached by these systems when they are
submitted to a periodic non-relaxational perturbation that mimics two
different extensive tapping mechanisms.

\subsection{Tapping dynamics}

The tapping is modelled by cycles consisting in two dynamical steps:
an ``energy injection'' step (called a tap for short) followed by a
zero temperature relaxation until blocking occurs
\cite{mehtabarker,mehtabarkerII,tetris,Brey,bergmehta,Dean}. During
the tapping the spins are randomly updated according two distinct
ways:
\begin{enumerate}
  
\item{\bf Thermal tapping (T)} The system undergoes a Monte-Carlo
  sweep at temperature $T$ with transition matrix specified by
  Eq.~(\ref{W_S}) or~(\ref{W_A}) depending on the model.
  
\item{\bf Random tapping (R)} Each variables is flipped in parallel
  with probability $p \in (0,\,\frac{1}{2}]$, irrespective of the
  kinetic constraints.

\end{enumerate}
During the tapping dynamics the detailed balance is broken, and after
a long enough time the system is expected to reach a steady state
regime in which the energy injected into it is in average
equal to that dissipated in the zero temperature relaxation steps.
We also checked that the steady state is independent of the initial 
configuration (with the exception of the lowest energy configuration
for the thermal tapping). 
Both tapping mechanisms coincide in the ``infinite tapping limit''
where a tap consists in reinitialising completely the system in a
random configuration. The dynamics in this limit has been recently
solved by De Smedt et al.~\cite{SGL} in kinetic 1D models similar to
the ones we study here, finding results in full agreement with ours.
For an analytical approach to the thermal tapping dynamic of the 1D
Fredrickson-Andersen model see also ref.~\cite{Brey}.  Note that the
blocking condition, namely that zeros are isolated, is obviously
independent of the tapping mechanism. However, the statistical
properties of the blocked configurations in the stationary state might
be (and actually are, as we will see) dependent on the way they are
typically accessed.
The set of blocked configurations for the model S and A is the same,
and one can compute their number ${\cal N}(e)$ as a function of the
energy density $e$, through simple combinatorial
arguments~\cite{CrRiRoSe}. In the thermodynamic limit this number is
exponentially large, and the Edwards entropy, which is by definition
$s(e)= \frac{1}{N} \log {\cal N}(e)$, reads:
\begin{equation}
  s(e) = e\log \frac{1+2e}{e} +(1+e)\log \frac{-1-2e}{1+e} \,,
  \label{s}
\end{equation}
from which one gets the inverse temperature or ``compactivity'' 
\begin{equation}
  \beta(e) \equiv \frac{\partial s}{\partial e}= \log
  \frac{(1+2e)^2}{-e(1+e)} \,.
  \label{beta}
\end{equation}
\begin{figure}
  \epsfxsize=3.4in
  \centerline{\epsffile{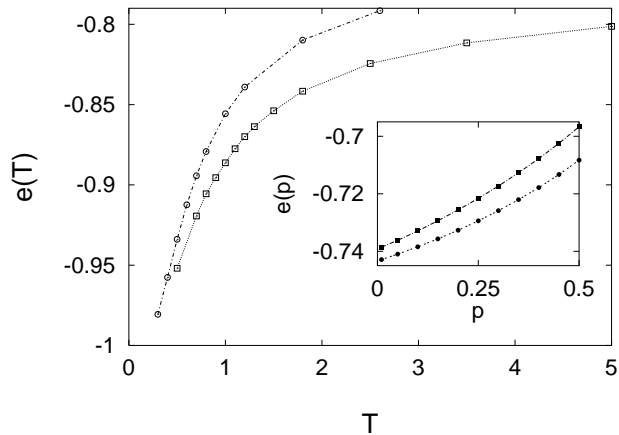}}
  \caption{Energy density vs tapping amplitude in the stationary state 
    of the tapping dynamics of the model S (circle) and model A
    (square).  Open symbols represents the {\sl thermal} tapping (main
    figure), while solid symbols correspond to the {\sl random}
    tapping (inset).}
  \label{E_multi}
\end{figure}  
We have performed extensive numerical simulations of both models A and
S with both thermal and random tapping dynamics.  We used spin chains
of length $N=2^{10},\,2^{15}$, checking finite-size effects against
$N=2^{20}$. The observable computed in the steady state regime were
typically averaged over samples of size $10^6,\,10^7$.  In
Fig.~\ref{E_multi} we plot the energy density in the stationary state
vs the tapping amplitude for the four possible cases we have examined.
We remark that curves corresponding to the two energy injection
mechanisms are rather different.  The random tapping explores only
configurations within quite a narrow interval of energy, and the zero
tapping limit of the steady state energy seems to converge to a value
higher than the ground state, where the blocked state entropy is still
extensive (see the inset of Fig.~\ref{E_multi}).  Decreasing $p$ below
the value $0.1$ does not yield substantially lower energies but only
makes longer the relaxation time to the stationary state. In the weak
tapping regime we have explored, $10^{-1} \ge p \ge 10^{-5}$, this
relaxation time goes like $\tau_{\rm \scriptstyle rel} \sim p^{-1}$.
With the thermal tapping mechanism instead, both models A and S are
able to explore a wider energy range and they appear to reach the
ground state as the tapping amplitude decreases to zero.  Blocked
configurations reached with random tapping are therefore less compact
of those reached with thermal tapping, leading for these models to a
non-universal (i.e.  dependent on the dynamical mechanism) asymptotic
packing density.  Also notice that with random tapping the asymptotic
energy density of the model A at a given $p$ is lower than the
corresponding one for the model S. This is easily understood as the
asymmetric constraint is stronger than the symmetric one and hence the
probability of the transition $0 \rightarrow 1$ in the zero
temperature relaxation step is higher in the model S than A.  For the
thermal tapping instead just the opposite happens.  In this case
during the energy injection step the spins can only be flipped by
respecting the kinetic constraints.  This gives a lower number of
spin-flip transitions in the model A with the respect to S (as the
latter is characterised by a weaker constraint), which eventually
results in a lower asymptotic energy for the model A.

The question that naturally arises in this context is whether models
with different energy vs tapping amplitude plots but with the same set
of blocked configurations may also share the same tapping
thermodynamics.  In order to investigate this point we measure several
observables in the stationary state of tapping dynamics and compare
their value with the corresponding observable analytically computed
with the Edwards measure.
\begin{figure}
  \epsfxsize=3.4in
  \centerline{\epsffile{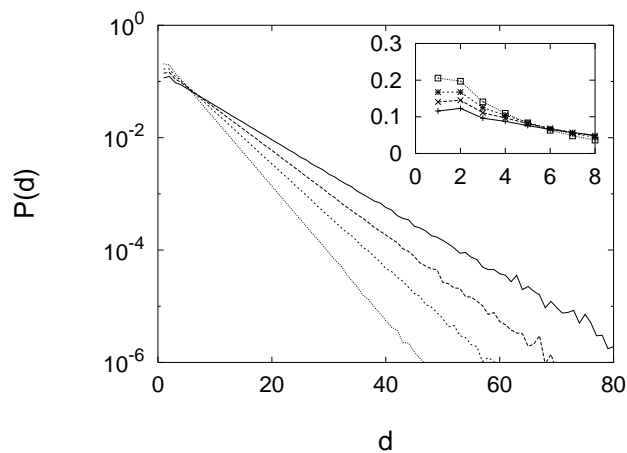}}
  \caption{Probability distribution of domain length, $P(d)$, 
    in the stationary state of the {\sl thermal} tapping dynamics of
    model A. Tapping temperature $T=1.0$, 1.3, 1.8, and 3.5.  The
    distribution is exponential for long domain size, but deviations
    (stronger the higher the tapping amplitude) from the pure
    exponential can be detected for short domains (see inset). Similar
    results were obtained for the {\sl random} tapping and for the
    Ising chain with symmetric constraint (model S).}
  \label{Pd_asym}
\end{figure}    

We first examine the probability distribution of domains size.  A
domain of size $d$ is defined here as a sequence of $d$ ones enclosed
by zeros.  The explicit computation reveals that within the uniform
measure, the distribution is exponential:
\begin{equation}
  P(d) = \frac{1+e}{-e} \left( \frac{2e+1}{e} \right)^{d-1} , \,\,\,\, 
  d \geq 1 \,.
  \label{Pd}
\end{equation}
Notice that this exponential distribution corresponds to independently
``throwing'' the number of zeros in a given interval, compatibly with
the blocking conditions and with no further correlations. Any
deviation from the exponential on the contrary implies correlations
induced by the dynamics, and our test can be seen as a measure of
these correlations.  In Fig.~\ref{Pd_asym} we show the function $P(d)$
as obtained from the thermal tapping of the A model.  We see that
while the exponential distributions works excellently for long domain
size, deviations are detected for short domains as a consequence of
the short-range correlations created by the kinetic constraints.
Similar results are obtained for the model S and in both models, more
pronounced deviations (not shown) are found for random tapping
dynamics.  In order to get a quantitative estimate of deviations we
measure the mean-squared fluctuations of domain length $\sigma_d^2$,
which in our case is given by
\begin{equation} 
  \sigma_d^2 = \frac{e(1+2e)}{1+e} \,.
  \label{d_dev}
\end{equation}
\begin{figure}
  \epsfxsize=3.4in \centerline{\epsffile{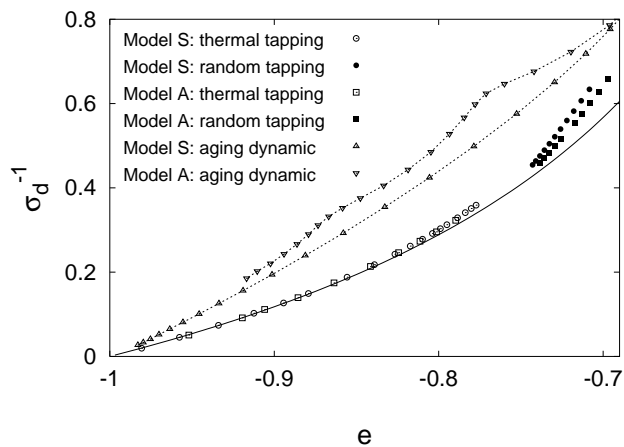}}
  \caption{Inverse of the mean squared fluctuation length $\sigma_d^{-1}$, vs 
    energy density $e$.  Circle (model S) and square (model A)
    represent numerical results for the stationary state of the
    thermal (open symbols) and random (solid symbols) tapping dynamics
    in the constrained Ising chains.  The full line represents the
    analytical result obtained from Eq.~(\ref{d_dev}). For comparison
    also shown are the numerical results for the non-stationary
    relaxational dynamics after a quench at temperature $T=0.2$ (model
    S, upward triangle) and $T=0.25$ (model A, downward triangle).}
  \label{fig:d_dev}
\end{figure}    
Notice that the average domain length $\overline{d}$,
\begin{equation} 
  \overline{d}  = \frac{-e}{1+e} \,,
  \label{d_ave}
\end{equation}
is not a good observable since it is only determined by the blocking
condition as a function of the energy, whatever the domain length
probability distribution.  In Fig.~\ref{fig:d_dev} we show
$\sigma_d^{-1}$ vs $e$ for both models and both tapping mechanisms.
We find that the flatness assumption over blocked configurations with
fixed energy works well for thermal tapping at low energy and does not
depend on the nature of kinetic constraints; while small but
systematic deviations are found at increasing energy.  For the random
tapping instead the deviations are found to be quite large at any
tapping intensity, showing that the sampling of configurations is not
ergodic with this energy injection mechanism.  One can easily check
that the energy interval explored with the random tapping dynamics
corresponds to a small region around the maximum of the entropy of
blocked configurations Eq.~(\ref{s}), where $e(s_{\rm \scriptstyle
  max}) \simeq -0.7236$. In this region the compactivity
Eq.~(\ref{beta}), is very small and the Edwards hypothesis is not
expected to hold~\cite{Edo,BaKuLoSe}.

For comparison we have also studied the aging dynamics, i.e. thermal
relaxation at a low temperature $T$ starting from a high energy random
configuration.  Given the simple one-dimensional nature of the model,
the system eventually equilibrates to the canonical distribution.
However, before equilibrium is reached, the system enters a scaling
regime during which the average domain length grows as
$\overline{d}(t) \sim t^{a T}$ for the model A~\cite{SoEv}, and in a
purely diffusive way, $\overline{d}(t) \sim
t^{1/2}$~\cite{Mauch_asym,CrRiRoSe}, in the model S.  In this regime
the domain sizes probability distribution is not exponential, and
although the average domain length closely approaches
Eq.~(\ref{d_ave}) (see the second of refs.~\cite{CrRiRoSe}), the
inverse of the mean squared fluctuation length remains far from the
Edwards value in the whole scaling regime for both models, except at
exceedingly low energy (see, Fig.~\ref{fig:d_dev}).

\begin{figure}
  \epsfxsize=3.4in
  \centerline{\epsffile{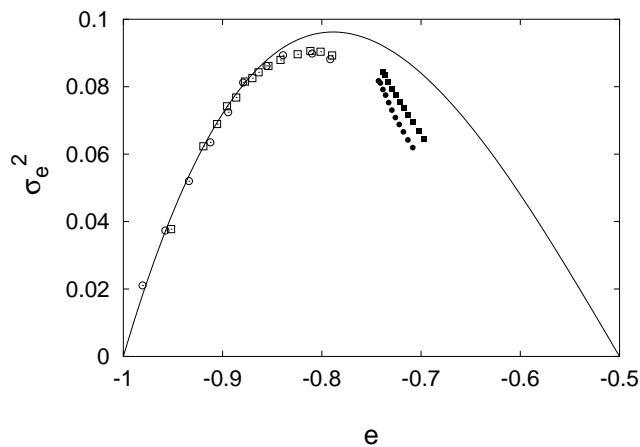}}
  \caption{Test of the fluctuation-dissipation relation Eq.~(\ref{fdt}),
    in the stationary state of the tapping dynamics of Ising chains
    with kinetic constraints.  Symbols correspond to the mean-squared
    energy fluctuations $\sigma_e^2$ vs energy density $e$,
    numerically obtained with the {\sl thermal} (open symbol) and the
    {\sl random} (solid symbol) tapping dynamics for the model S
    (circle) and model A (square).  The full line represents the
    analytic result~Eq.~(\ref{sigma_e}).  }
  \label{fdt_tapping}
\end{figure} 

Another aspect of the Edwards thermodynamic construction  concerns 
the behaviour of the energy fluctuations.  By standard thermodynamics,
in the regimes well approximated by the Edwards hypothesis, the {\it
  spatial} fluctuations of the energy $\sigma_e^2$, should follow the
canonical relation
\begin{equation}
{\sigma_e^2} = - \frac{\partial e}{\partial \beta} \,, 
\label{fdt}
\end{equation}
which in our case gives 
\begin{equation}
\sigma_e^2 = e(1+e)(1+2e)    \,.
\label{sigma_e}
\end{equation}
One may wonder whether in that regime the {\it temporal} fluctuations
follow the same law, hinting for a canonical distribution of the
blocked states.  We find that the energy fluctuations in the
stationary state of the tapping dynamics are essentially Gaussian
distributed.  In Fig.~\ref{fdt_tapping} we compare the mean-squared
fluctuations of energy $\sigma_e^2$, with the analytic result.  We see
that in the region where the uniform hypothesis works well, the
fluctuations follow within numerical error the law implied by the
canonical statistics, with improving accuracy as the tapping intensity
is decreased. Consistently with our previous results we find that for
random tapping the Eq.~(\ref{sigma_e}) is violated.

Small but systematic deviations at higher tapping intensity can also
be observed in the random tapping dynamics of a one dimensional Ising
chain recently studied by Dean and Lef\`evre~\cite{LeDe}.
Interestingly, we have found that the {\sl thermal} tapping dynamics
of this model gives results similar to the random tapping.  Hence, at
variance with the kinetically constrained Ising chains, in the
Dean-Lef\`evre model there is substantially no difference between the
two distinct energy injection mechanisms.

In conclusion, our results show that for thermal tapping the Edwards
measure is a good approximation independently of the kinetic
constrains, but it cannot be considered exact as systematic deviations
appear at increasing tapping intensity.  Random tapping instead
prevents the system from reaching high compactivity states.  This
results in a disagreement with Edwards measure, showing that the
flatness assumption can be very sensitive to the nature of the energy
injection mechanism and its interplay with blocking condition.

\section{The three-spin model on random graphs}
\label{johannes}

In this section we focus on the results of random and thermal tapping
applied to three-spin models defined on a random graph with fixed
connectivity $k$. The model is defined by the Hamiltonian
\begin{equation}
  \label{hdef}
  H=-\sum_{l<m<n} C_{lmn} S_l S_m S_n 
\end{equation}
where the connectivities $C_{lmn}$ are invariant under permutation of
the indices and are chosen at random with the constraint that
$\sum_{m<n} C_{lmn}=k \, \forall l$. The three-spin model (albeit with
fluctuating connectivities) has been used to model granular compaction
\cite{bergmehta}, since it features states with locally minimal
energies (satisfied plaquettes are +++,--+, and permutations of the
latter), which however may be globally incompatible with one another.
Under the term 'geometric frustration' the same mechanism is thought
to be at the heart of the slow compaction of granulars.  The model is
also attractive as the aging evolution should obey mean field theory,
where the asymptotic validity of flat measure on
Thouless-Anderson-Palmer states is well established, and one can test
whether tapping and glassy relaxational dynamics are by some means
related~\cite{stadlerbergmehta}. However the fact that tapping is
non-thermal complicates the issue - for instance for large tapping
amplitudes one cannot expect a flat measure over blocked states to
hold.  Consider as an extreme case random tapping with $p=1/2$, which
corresponds to a series of quenches from random initial
conditions. Since the typical blocked configurations do not - in
general - have the largest basin of attraction among each other, we
would not expect the flat measure over the blocked states to hold.

Configurations are deemed blocked if $h_l s_l \geq 0$ where
$h_l=\sum_{m<n} C_{lmn} s_m s_n$ is the local field acting on each
site.  In order to check the Edwards hypothesis for such models, the
statistical mechanics of blocked configurations of these models must
be worked out. In principle, this would involve averaging the
logarithm of the number of blocked configurations over the disorder,
i.e.  the different graphs (quenched average).  From an analytical
point of view, fixed connectivity graphs provide a simple
testing-ground, since using the methods of
\cite{bergsellitto,franzetal} one finds that for sufficiently high
energies the annealed average of the number of blocked configurations
gives the same result as the quenched average.

The number of blocked configurations ${\cal N}(e)$ at a given energy
density $e$ may be written easily as
\begin{eqnarray}
\label{nblocked}
{\cal N}(e) & = & \prod_l \left[ \sum_{s_l=\pm 1} \sum_{h_l=-\infty}^{\infty} 
        \delta \left( h_l-\frac{1}{2} \sum_{m,n} C_{lmn}s_m s_n \right) 
\right. \nonumber \\	
&&   \left.      \times \Theta \left( h_l s_l \right) \right]  
     \delta \left( e - \frac{1}{3N}\sum_l h_l s_l \right) \ ,
\end{eqnarray}  
where $\delta(x)$, denotes a Kronecker-delta $\Theta(x)$ denotes a
Heaviside step-function with $\Theta(x)=1 \ \mbox{if} \ x \geq 0$ and
$0$ otherwise, and ${\rm e}^x$ denotes the standard exponential function.
After using integral representations for the Kronecker-deltas and
standard manipulations \cite{deanold}, one obtains the entropy of
blocked states in the annealed approximation
\begin{eqnarray}
\label{sbannres}
 s(e) &=& \frac{1}{N}\ln \langle\!\langle {\cal N}(e) 
		\rangle\!\rangle = \mbox{extr}_{a,b,\beta} 
	\left[ \beta e - \frac{8}{3} \ (a^3+ b^3) + 
  \frac{2}{3} k (1- \ln k)  \right. \nonumber \\
&&  + \left. \ln \left(  (2ab)^k \sum_{h=0(1)}^{\prime k} 
    ({\rm e}^{\beta/3} \frac{a}{b})^{h} 
    \left( {k \choose \frac{k-h}{2}} + {k \choose \frac{k+h}{2}} \right)
  \right) \right] \, 
\end{eqnarray} 
where the angular brackets denote the average over graphs of fixed
connectivity $k$ and $a,b,\beta$ are to be determined by extremising
this expression with respect to these three parameters. The sum over
$h$ proceeds in steps of two, as for even $k$ thus only even local
fields are possible, and likewise for odd values of $k$.

Having solved the self-consistent equations for the three parameters,
one may also determine the fraction of sites $g_h$ with a given value
of the local field $h$ (even $h$ for even $k$ and odd $h$ for odd $k$)
\begin{equation}
\label{gh}
g_h = \frac{ 
  ( {\rm e}^{\beta/3} a/b )^{|h|}  {k \choose \frac{k-h}{2}}
    \left[ 1 + \delta(h) \right]  }{ 
    \sum_{h=0(1)}^{\prime k} \left( {\rm e}^{\beta/3} a/b \right)^h 
    \left[ {k \choose \frac{k-h}{2}} + {k \choose \frac{k+h}{2}} \right] 
      } \ .
\end{equation} 
For $k>3$ the fraction of sites with a certain local field serves as a
convenient test of the Edwards hypothesis by comparing the values of
$g_h$ reached asymptotically with those predicted by the flat measure
over all blocked configurations at the asymptotic energy.  (For $k=3$
and for a symmetric distribution of the local fields, the fraction of
sites with $h=1$ versus those of $h=3$ is a unique function of the
energy in all configurations).

\begin{figure}  
  \epsfxsize=3.in
  \centerline{\epsffile{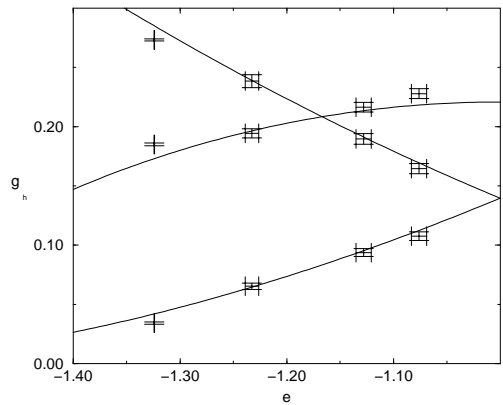}}
  \caption{Asymptotic results for connectivity $k=5$ and {\it thermal} 
    tapping with $T=.5,2,2.86,5$ (from left to right). The solid lines
    give the corresponding fractions of sites with $h_i=1,3,5$
    according to the flat measure (bottom to top on the lhs).}
  \label{figasymtk5b}
\end{figure}

\begin{figure}
  \epsfxsize=3.in
  \centerline{\epsffile{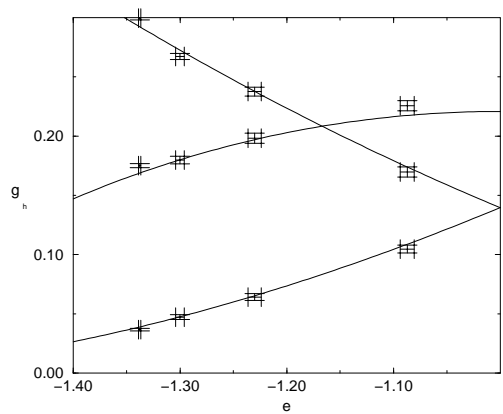}}
  \caption{Asymptotic results for connectivity $k=5$ and {\it random} tapping 
    with $pN=300,500,1000,2000$ (lower tapping amplitudes did not
    yield substantially lower energies). The solid lines give the
    corresponding fractions of sites with $h_i=1,3,5$ according to the
    flat measure (bottom to top on the lhs).  }
  \label{figasymtk5r}
\end{figure}

In the following we compare the results of thermal and of random
tapping for $k=5$ and $k=6$. We use system sizes of $N=10^4$. An
asymptotic state was typically reached after $10^6$ taps in all but
the lowest intensities, where up to $3 \times 10^6$ taps were
necessary.  Since graphs of fixed connectivity are highly homogeneous
no sample averaging was necessary. In Figs.~\ref{figasymtk5b}
and~\ref{figasymtk5r} the asymptotic results averaged over $1000$
steps with the errorbars giving the standard deviation of the energy
and the fraction of sites with a given local field also measured over
$1000$ steps. We plot the fraction of sites of a given local field
against the asymptotic energy and compare the results with the
prediction of (\ref{gh}).

As expected there are discrepancies between the numerical results of
the tapping dynamics and the analytical results for the flat measure
at high amplitudes of both thermal and random tapping. However also at
low amplitudes small discrepancies are found. These are probably due to a
dynamical slowing down and diverging equilibration times at low
temperatures. 

\begin{figure}
  \epsfxsize=4.3 in
  \centerline{\epsffile{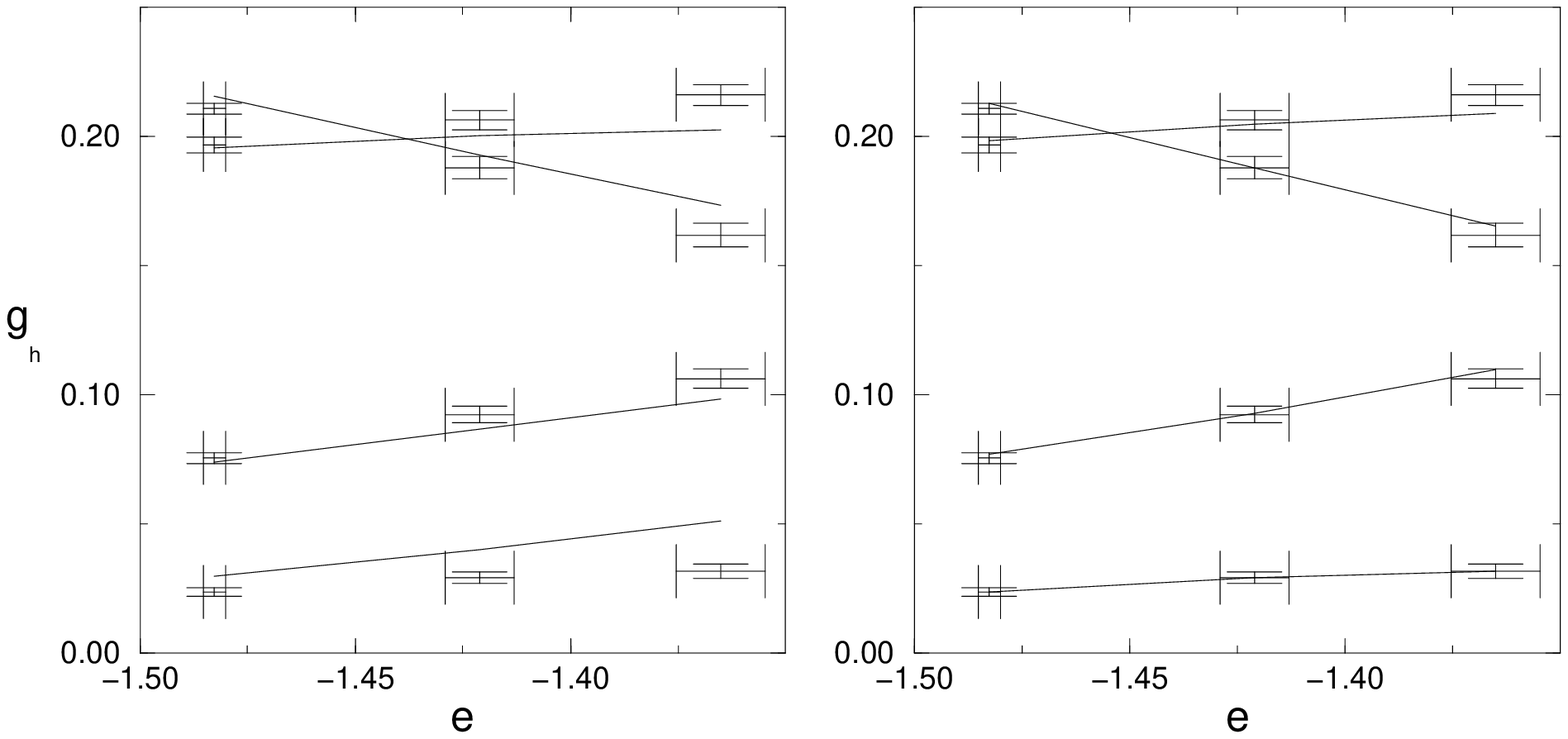}}
  \caption{Asymptotic results for connectivity $k=6$ and {\it thermal}
    tapping with $T=2,2.86,5$ (from left to right, lower amplitudes
    did not yield substantially lower energies), plotting the
    fractions of sites with $h_i=0,2,4,6$ (bottom to top on the lhs)
    against the energy. In the left graph, the solid lines give the
    corresponding analytic result according to the uniform measure. In
    the graph on the right hand side, the same numerical results are
    plotted and compared to the restricted measure plotted as solid
    lines.  }
  \label{figasymtk6b}
\end{figure}

\begin{figure}
  \epsfxsize=4.in
  \centerline{\epsffile{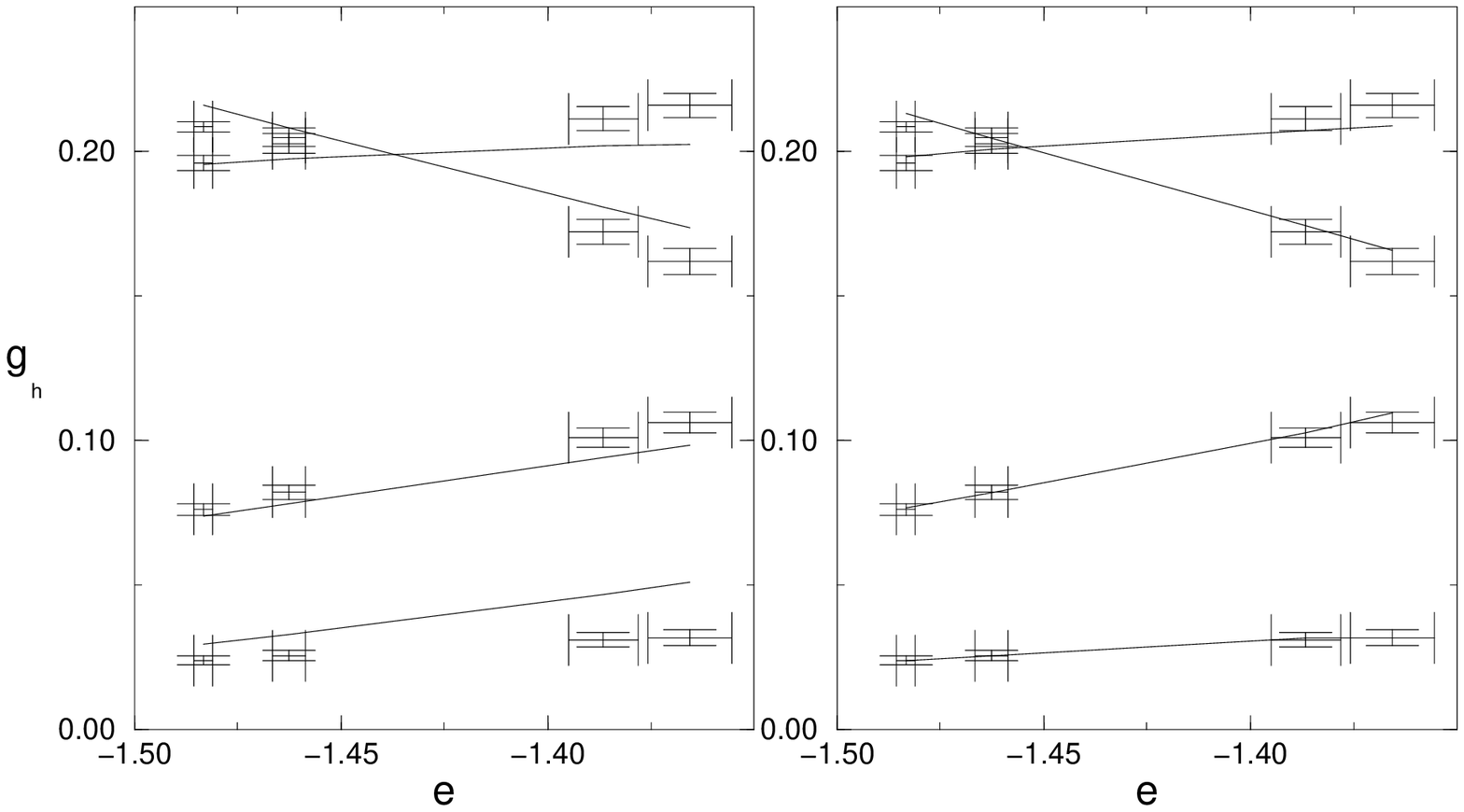}}
  \caption{Asymptotic results for connectivity $k=6$ and {\it random} 
    tapping with $pN=300,500,1000,2000$ (lower tapping amplitudes did
    not yield substantially lower energies), again plotting 
    the fractions of sites with $h_i=0,2,4,6$ (bottom to top on the lhs)
    against the energy. In the left graph, 
    the solid lines give the corresponding analytic result according to the
    uniform measure. In the graph on the right hand side, the same results 
    numerical results are plotted and compared to the restricted measure 
    plotted as solid lines.  }
  \label{figasymtk6r}
\end{figure}

The situation is however more drastic in the case of $k=6$, shown in
Figs.~\ref{figasymtk6b} and~\ref{figasymtk6r} where both for thermal 
and for random tapping respectively the flat measure does not agree
with the numerical results at any tapping amplitude.


This effect is due to a simple but crucial difference between graphs of 
odd and even constant connectivity: 
Graphs with even connectivity qualitatively differ from those with odd 
connectivity since in the former case sites may have their local magnetic 
field equal to zero so their spins
may be free to flip without changing the energy. The dynamics of these
spins is crucial at low temperatures as they correspond to neutral
directions in phase space~\cite{barratzecchina}. 

In odd-connectivity graphs such neutral directions are absent, whereas 
in generic graphs with fluctuating connectivity as considered in 
\cite{bergmehta} they are also present. 

The spins with zero magnetic field may be thought of as being exposed 
to a continuous tapping process even during the quench phase.  
The idea that this process distorts the flat measure may be used to promote
the fraction of spins with zero local field to a relevant macroscopic
variable, like the energy. In fact, a better comparison with the
dynamics is obtained considering the ensemble of all blocked
configurations of a given energy {\it and} of a given fraction $g_0$
of sites with zero local field. The calculation is a simple variation
on (\ref{sbannres}) and yields
\begin{equation}
\label{gh2}
g_h = \frac{ 
  ( {\rm e}^{\beta/3} a/b )^{|h|}  {k \choose \frac{k-h}{2}}
    \left[ 1 + \delta(h) \right]{\rm e} ^{ - \hat{g}_0 \delta_{h,0} }  }{ 
    \sum_{h=0(1)}^{\prime k}{\rm e}^{ - \hat{g}_0 \delta_{h,0} } 
	\left( {\rm e}^{\beta/3} a/b \right)^h 
    \left[ {k \choose \frac{k-h}{2}} + {k \choose \frac{k+h}{2}} \right] 
      } \ .
\end{equation} 
where the order parameters are determined by the extremal condition in 
\begin{eqnarray}
\label{sbannres2}
&& s(e) =  \mbox{extr}_{a,b,\hat{g}_0,\beta} \left[ \hat{g}_0 g_0 + 
  \beta e - \frac{8}{3} \ (a^3+ b^3)  + \frac{2}{3} k (1- \ln k)
	+ \right. \nonumber \\
&  & \left.  \ln \left( (2ab)^k \sum_{h=0(1)}^{\prime k} {\rm e}^{-\hat{g}_0 \delta_{h,0}} 
	\left({\rm e}^{\beta/3}\frac{a}{b}\right)^h  
	\times \left({k\choose\frac{k-h}{2}}+{k\choose\frac{k+h}{2}}\right) 
	\right) \right] \ . 	
\end{eqnarray} 

To test this new ensemble we compare the values
of $g_2,g_4,g_6$ at the asymptotic state with those given analytically
by the restricted ensemble at the asymptotic values both of the energy
and of $g_0$. The results are shown in Figs.~\ref{figasymtk6b}
and~\ref{figasymtk6r}. Except at high amplitudes, the numerical and
analytical results agree very well. Clearly more information on the blocked 
states is used in the restricted measure, so some improvement of the fit 
between analytical and numerical results is expected solely on these grounds. 
Similarly the agreement in the case of the fraction of sites with zero local 
field $g_0$ is solely due to the fitting. Nevertheless the fact that the flat 
measure fails in the case of even-connectivity graphs shows the crucial 
role of sites with zero magnetic field. The restricted measure is the 
simplest way of modifying the flat measure in this case. 

Of course the relevance of such
neutral moves to realistic models of e.g.  of granular is debatable,
however loosely constrained particles termed 'rattlers' have been
found to cause subtle dynamical effects in simulations of granular
particles \cite{rattlers}.

\section{Summary and Conclusions}

In this paper we have studied stationary dynamical measures of several
abstract spin models endowed with two kinds of energy injection
mechanism: random and thermal tapping.  We first considered two
kinetically constrained Ising chains (with symmetric and asymmetric
constraints) having the same entropy of blocked states.  We find that,
in the case of the thermal tapping, the Edwards measure gives a good
approximation for the observables we studied, {\sl independently} of
the kinetic constraints.  This can be understood as the uniform
measure implies uncorrelated domains of up spins, and the dynamics
does not create spatial long range correlations. Despite that, small
correlations are always dynamically induced and systematic deviations
are found. As one could expect, the quality of the approximation
improves as one goes towards lower tapping amplitudes and lower
energies. Deviations are particularly evident in the case of random
tapping showing that the energy injection mechanism may have a strong
influence on the nature of the asymptotic stationary regime.  We
interpret these deviations as essentially due to the vanishing
compactivity of blocked configurations reached by random tapping.  We
finally observed that non-ergodic sampling occurs during the aging as
the purely relaxational dynamics of these Ising chains is a domain
growth process.  Nevertheless, this does not prevent the Edwards
measure to be a good approximation for the steady state of the thermal
tapping. This suggests that ergodicity in the stationary regime
generally requires less stringent conditions than the aging dynamics.

Similar results are obtained in the case of the 3-spin model on the
random hypergraph, where the Edwards measure gives a reasonable
approximation both for thermal and for random tapping provided there
are no neutral directions in phase space.  In the latter case,
realized by sites with zero local field, we introduced a restricted
measure of blocked configurations with a {\it given} fraction of sites
with zero local fields.

It is an open problem if there are systems where the uniform
measure on blocked states is exact for tapping dynamics. In this paper
we showed that at least for thermal tapping, where the energy
injection step is correlated with the energy landscape, the Edwards
hypothesis is a good approximation whose quality increases with
decreasing tapping amplitude.

\section*{Acknowledgements}

S.F. acknowledges interesting discussions with A. Barrat and
J.M. Luck. We thank G. De Smedt, C. Godreche and J.M. Luck for
communicating us their results before publication.

\section*{References}

\end{document}